\documentclass[draft,jgrga]{AGUTeX}

\usepackage{lineno}

\linenumbers*[1]

\usepackage[dvips]{graphicx}

\authorrunninghead{SUBRAMANIAN \& CAIRNS}

\titlerunninghead{CONSTRAINING CORONAL TURBULENCE WITH RADIO DATA}

\authoraddr{Prasad Subramanian, Indian Institute of Science Education and Research, Sai Trinity Building, Pashan, Pune - 411021, India.
(p.subramanian@iiserpune.ac.in)}
\authoraddr{Iver Cairns, School of Physics, University of Sydney, NSW 2006, Australia. (cairns@physics.usyd.edu.au)}

\begin{document}


%


%


\title{Constraints on coronal turbulence models from source sizes of noise storms at 327 MHz}

%





%


%


















%


%




\begin{abstract}
We seek to reconcile observations of small source sizes in the solar 
corona at 327 MHz with predictions of scattering models that
incorporate refractive index effects, inner scale effects and a 
spherically diverging wavefront. We use an empirical prescription 
for the turbulence amplitude $C_{N}^{2}(R)$ based on VLBI observations 
by Spangler and coworkers of compact radio sources against the 
solar wind for heliocentric distances $R \approx$ 10--50 $R_{\odot}$. 
We use the  Coles \& Harmon model for the inner scale $l_{i}(R)$, that is 
presumed to arise from cyclotron damping. In view of the prevalent 
uncertainty in the power law index that characterizes solar wind turbulence 
at various heliocentric distances, we retain this index as a free parameter.
We find that the inclusion of spherical divergence effects suppresses the 
predicted source size substantially. We also find that inner scale effects 
significantly reduce the predicted source size. An important general 
finding for solar sources is that the calculations substantially underpredict the observed 
source size. Three possible, non-exclusive, interpretations of this 
general result are proposed. First and simplest, future observations 
with better angular resolution will detect much 
smaller sources. Consistent with this, previous observations of small 
sources in the corona at metric wavelengths are limited by the instrument resolution. Second, 
the spatially-varying level of turbulence $C_{N}^{2}(R)$ is much 
larger in the inner corona than predicted by straightforward extrapolation 
Sunwards of the empirical prescription, which was based on observations between 10--50 $R_{\odot}$. Either the functional form or the 
constant of proportionality could be different. Third, perhaps the inner 
scale is smaller than the model, leading to increased scattering. 

\end{abstract}


%


%



%



\begin{article}


%


%


\section{Introduction}
Refractive scattering of radiation by density turbulence in the Sun's corona 
and solar wind leads to angular broadening of embedded radio sources, and of cosmic 
sources observed through these media. The process is similar to the twinkling of 
stars and modified``seeing'' caused by density turbulence in Earth's atmosphere 
and ionosphere. This scattering process has been investigated for many years 
using geometrical optics [e.g., Steinberg et al 1971] and the parabolic wave 
equation [e.g., Lee and Jokipii, 1975; Coles and Harmon 1989; Bastian 1994; Cairns 1998]. 

Scattering is thought to affect the observed properties of type II and III solar 
radio bursts in several ways: greatly increasing the angular sizes of the sources 
[e.g., Riddle 1974], causing the time profiles to have exponential decreases 
[e.g., Robinson and Cairns, 1998], and causing anomalously low brightness 
temperatures at decametric wavelengths [e.g., Thejappa \& Kundu 1992]. 

The primary motivation of this paper is to investigate the constraints imposed on 
models of density turbulence in the solar corona
by recent observations at 327 MHz [Mercier et al., 2006], made by combining visibilities from the Giant 
Metrewave Radio Telescope (GMRT) in Pune, India, and the Nancay 
Radioheliograph (NRH) in France. 
The maps of Mercier et al. [2006] show structures ranging from the 
smallest observed size of $49^{''}$ to that of the whole Sun, with 
dynamic ranges as high as a few hundred. These features make them 
the best meter wavelength snapshot maps of the solar corona to date.
Mercier et al. [2006] found 
the smallest steady angular size of type I solar noise storms to be 
$49^{''}$ in their high dynamic range, full disk, 17 second snapshots. We therefore adopt the smallest observed source size of $49^{''}$ 
in these maps to be a canonical number for comparison with our model 
predictions. The only other two-dimensional map showing small source 
sizes that we are aware of is that of Zlobec et al [1992], who 
observed a source as small as 
$30^{''}$ at 327 MHz. However, the dynamic range of their map was 
severely limited, and included only a very small range of size 
scales. It should be noted that the lower limit to the
observed size is imposed by the resolution of the instrument if 
scattering by density turbulence is weak enough. The scattering calculations shown
below imply that there is a possibility that smaller solar sources will be 
detected in the future by instruments with improved angular resolution. 

In this paper we use a formalism based on the paraxial wave equation and the 
structure function, together with observationally based models for the 
density turbulence that scatter the radiation, to 
predict the size of sources in the solar corona at 327 MHz. 
Our results can be interpreted as the scatter-broadened 
image of an ideal point source in the solar corona. 
We have used an empirical model for the amplitude $C_{N}^{2}(R)$ 
of coronal turbulence that is based directly a fit to the scattering 
measure obtained from VLBI observations of cosmic sources 
broadened by scattering in the outer solar corona and inner solar wind. 
Here $R$ is the heliocentric distance. We have assumed that this model 
is valid throughout the corona, specifically at smaller $R$. 
We also consider the effects of spherical and plane wave 
propagation, variations of the inner scale $l_{i}(R)$ and power-law 
index $\alpha$ of the turbulence on the predicted source sizes. In most 
cases, we find that the models predict sizes that are at least an 
order of magnitude below the smallest observed size of 49$^{''}$ at 327 MHz.  
Our formalism and analyses differ primarily from those of 
Bastian [1994] in the models for $C_{N}^{2}(R)$ and the electron 
density profile $n_{e}(R)$, 
while our applications are to metric rather than centimetric and 
decimetric emissions. Since our predictions are much smaller and 
Bastian's [1994] predictions much larger than $49''$ at $327$ MHz, 
the analyses demonstrate the importance of knowing $C_{N}^{2}(R)$, $n_{e}(R)$, 
and $l_{i}(R)$  much better for future observations and predictions of 
solar sources. These quantities are also relevant to the heating 
and outward flow of the coronal plasma, with activity localized 
to specific ranges of $R$ potentially leading to larger $C_{N}^{2}(R)$ 
and so enhanced scattering at, say, decimetric frequencies 
than expected at, say, metric frequencies.

The paper is organized as follows. In \S~2 we summarize 
the scattering formalism and observations of the density turbulence.  
Coronal density models are described in \S~3. The results are 
presented in \S~4, including estimates of the predicted angular 
broadening, and the implications for coronal density turbulence. 
The conclusions are presented in \S~5.

\section{Angular broadening}

We first consider the angular broadening predicted by the empirical formula 
of Erickson (1964):
\begin{equation}
\theta = 50 \left( \frac{\lambda}{D} \right)^2 \mbox{ arcminutes} \ .
\label{eq0}
\end{equation}
Here $\lambda$ is the observing (free-space) wavelength in meters and 
$D$ is the elongation in units of $R_{\odot}$. If we take 
$D = 1.056 R_{\odot}$, which is where 327 MHz emission would originate 
according to the hybrid density model described below, then
$\theta \simeq 50^{'}$ for 327 MHz, which is much larger (60 times) than 
the observed $49^{''}$. It points to a significant difference between 
the situation for observations of celestial background sources against 
the solar wind, for which Erickson's (1964) formula is well accepted, 
and observations of solar radio events that originate in the solar 
corona. This difference will be specifically addressed in the Discussion 
section below.

\subsection{Density turbulence}

Density turbulence in the Sun's corona and solar wind is modeled here by 
writing the three-dimensional isotropic spatial power spectrum 
$S_{n}(k,R)$ of the fluctuating part 
$\delta n$ of the electron density $n_{e}$ as 
[cf., Lee \& Jokipii 1975; Rickett 1977; Coles and Harmon, 1989; Bastian, 1994; 
Cairns 1998; Spangler, 2002]  
\begin{equation}
S_{n}(k,R) =  \langle (\delta n)^{2} \rangle (k,R) = C_{N}^{2}(R) k^{-\alpha} e^{-k^{2}/q_{i}^{2}} \ . 
\end{equation}
Here $k$ and $R$ are the (isotropic) wavenumber and radial distance (in units of $R_{\odot}$), 
respectively, $C_{N}^{2}(R)$ 
models the level of turbulence, $\alpha$ is the power-law index, and 
$q_{i}$ is the wavenumber corresponding to the inner scale of the turbulence. 
While it is fairly well established that the turbulence spectrum largely follows the
Kolmogorov scaling (with $\alpha = 11/3$) at scales larger than about $100$ km, there is some
evidence that it flattens, with $\alpha$ decreasing to values as low as $3$, at scales
between a few km and a few hundred km [Bastian, 1994]. There is also some 
evidence for variation of the turbulence power law spectrum with heliocentric 
distance [e.g., Efimov et al., 2008]. Furthermore, there is evidence for significant variation
in the index between the slow and fast solar wind [Manoharan et al., 1994]. 
We therefore retain $\alpha$ as a parameter. It may be noted that some authors 
use a power law index of $5/3$ to describe the one-dimensional Kolmogorov 
spectrum; the index they refer to is equal to $\alpha - 2$.

The empirical model we use for $C_{N}^{2}(R)$ was originally mooted by 
Armstrong \& Woo [1980] and later refined, based on VLBI 
observations between 10--50 $R_{\odot}$, by Spangler \& Sakurai 
[1995] and Spangler [2002] among others: 
\begin{equation}
C_{N}^{2}(R) = 1.8 \times 10^{10}\,\biggl ( \frac{R}{10 R_{\odot}} \biggr )^{-3.66}\,\,\,\, \ .  
\label{eq3}
\end{equation}
The dimensions of $C_{N}^{2}$ depend on $\alpha$, being m$^{-\alpha - 3}$. The normalizations for $C_{N}^{2}$ differ by about a factor of a few and the power-law index 
with $R$ ranges from $-3.66$ to $-4$ in these works, presumably due to solar wind variability.  

The inner scale $l_{i}$ is modeled using Coles \& Harmon's [1989] 
model which agrees roughly with their observations, 
\begin{equation}
q_{i}(R) = \Omega_{i}(R)/3 V_{A}(R) \equiv 2 \pi/l_{i}(R) = \frac{2 \pi}{684 \, n_{e}(R)^{-1/2}}\,\,\,\,\,{\rm km}^{-1}
\label{eq2}
\end{equation}
where $\Omega_{i}$ is the ion cyclotron frequency, $V_{A}$ is the Alfv\'en speed and $n_{e}$ is the electron density in ${\rm cm}^{-3}$. This model is interpreted 
conventionally in terms of cyclotron damping by MHD waves. We use this definition for 
the inner scale throughout this paper, except in two cases where we artificially set 
$l_{i}$ equal to a very small value.

A popular alternative prescription for $C_{N}^{2}$ supposes that $C_{N}^{2}$ is $\propto$ the square of the background electron density. Such a prescription has a constant of proportionality, which is often determined via observed values of the phase structure function (e.g., Bastian 1994). The magnitude of the phase structure function in turn, is very dependent on the elongation to which it is referenced. We discuss this issue further in \S~4.

\subsection{Plane vs spherical wave propagation}

Scattering depends quantitatively on whether the wavefront is planar (1-D) or spherical (3-D). When a source is embedded in the scattering medium, it is often appropriate to adopt a formalism that includes the spherically diverging nature of the wavefront. The geometry for spherically diverging propagation is shown in \callout{Figure 1}. 

\begin{figure}
\includegraphics[width=40pc]{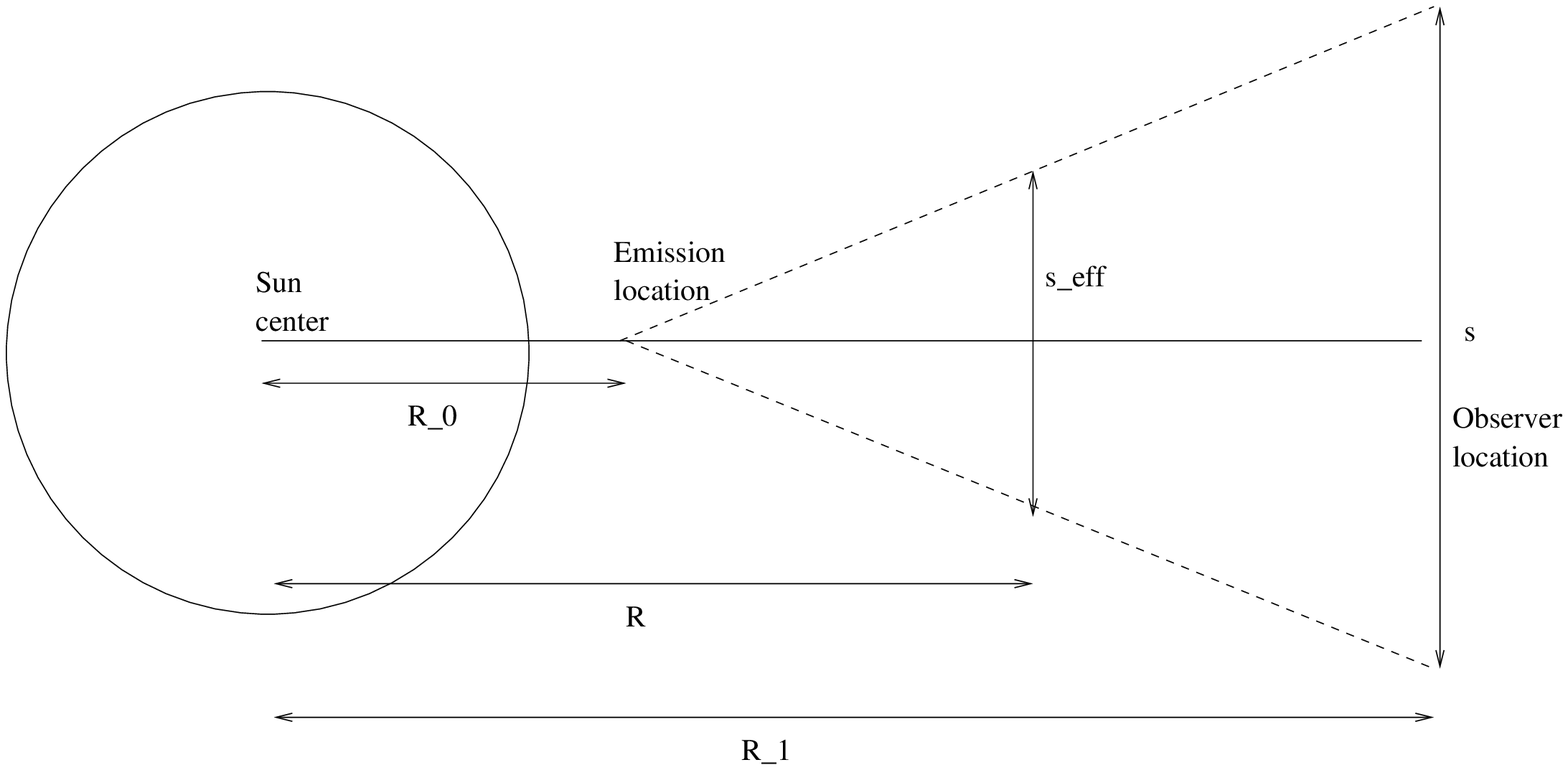}
\caption{Geometry for spherically diverging wavefront, which is 
appropriate to a situation where the source is embedded in the 
scattering medium}
\end{figure}

Similarly, when a plane wave illuminates the scattering medium, a 1-D planar formalism is standard. In this case an observer is typically sensitive only to scattering regions (eddies) with sizes of order the baseline length s. In the spherically diverging situation, however, the observer is sensitive to a range of eddy sizes given by $s a/b$, where $a$ is the (continuously varying) distance of the scattering screen from the source and $b$ is the distance of the observer from the source. In our situation, this is tantamount to saying that the effective baseline for spherical wave propagation is [Ishimaru, 1978]
\begin{equation}
s_{\rm eff} = s R/(R_{1} - R_{0}) \, .
\label{eq2a1}
\end{equation}
This is the basic difference between Eqs~(\ref{eqp2v43}) and (\ref{eqp2v44}) discussed below.

In the solar situation, radiation from an embedded coronal source 
is subject to scattering as it propagates to the observer. Since 
the radiation is generated near $f_{p}$ and 2$f_{p}$, scattering 
effects are expected to be largest in the source and in its vicinity, 
assuming that $f_{p}(R)$ decreases monotonically with increasing R. 
Spherical effects are expected to arise in two ways. Firstly, scattering 
will maximally distort an initially plane wavefront close to and in 
the source. Secondly, on a larger scale, the solar wind density 
is expected to be spherically symmetric, with radiation being 
refracted towards the radial direction. Accordingly, spherical 
divergence effects are expected to be vital. They are explicitly 
calculated below and shown to be quantitatively important. In 
contrast, the planar formalism is expected to be appropriate when 
the source, scattering region(s), and observer are all far apart, 
as assumed in calculations for pulsars and other celestial sources.

\subsection{Structure function}
The starting points for the expression we use for the scattering angle are equations (4)--(7) of Coles et al [1987] that specify the structure function and the mutual coherence function using the parabolic wave equation (PWE) formalism that includes small-angle refractive scattering 
and diffraction, but not reflection. For the sake of completeness, we reproduce them below. The asymptotic forms of the gradient of the phase structure function $D(s, R)$ are
\begin{equation}
\frac{\partial}{\partial R}\,D(s,R) = \frac{8 \pi^{2}}{2^{\alpha - 2}\,(\alpha - 2)}\, \frac{\Gamma ( 1 - (\alpha - 2)/2 )}{\Gamma ( 1 + (\alpha - 2)/2 )} \, C_{N}^{2}(R)\,r_{e}^{2}\,\lambda^{2}\,s^{\alpha - 2}\, , \,\,\,\,{\rm for}\,s_{\rm eff} \gg l_{i}(R)\, ,
\label{eqp2v41}
\end{equation}
\begin{equation}
\frac{\partial}{\partial R}\,D(s,R) = \frac{4 \pi^{2}}{2^{\alpha - 2}}\,\Gamma \biggl ( 1 - \frac{(\alpha - 2)}{2} \biggr ) \,C_{N}^{2}(R)\,r_{e}^{2}\,\lambda^{2}\,l_{i}(R)^{\alpha - 4}\,s^{2}\, , \,\,\,\,{\rm for}\,s_{\rm eff} \ll l_{i}(R)\, ,
\label{eqp2v42}
\end{equation}
where $r_{e}$ is the classical electron radius, $\lambda$ is the observing wavelength and $s_{\rm eff}$ is the effective interferometer spacing. It is noted that this formalism is valid only for $2 < \alpha < 4$; in particular, equations (\ref{eqp2v41}) and (\ref{eqp2v42}) diverge at $\alpha = 4$ owing to the behavior of the term $\Gamma ( 1 - (\alpha-2)/2)$. It may also be noted that the branches (\ref{eqp2v41}) and (\ref{eqp2v42}) do not meet at $s_{\rm eff} = l_{i}$; the ratio of (\ref{eqp2v42}) to (\ref{eqp2v41}) is equal to $(1/2) (\alpha - 2) \, (l_{i}/s)^{\alpha - 4}\, \Gamma ( 1 + (\alpha-2)/2)$, and at $s_{\rm eff} = l_{i}$ this is equal to unity only for $\alpha = 4$.

The effective interferometer spacing $s_{\rm eff}$ is equal to $s$ for the case of plane wave propagation, but is equal to $s R/(R_{1}-R_{0})$ for spherical wave propagation, as discussed in \S~2.2 and Figure 1.
The phase structure function for the cases of plane wave and spherical wave propagation are
\begin{equation}
D_{p}(s) = \int_{R_{0}}^{R_{1}} \frac{\partial}{\partial R}\,D(s,R)\,dR\, , \,\,\,\,{\rm for\,\,plane\,\,wave\,\,propagation}\, ,
\label{eqp2v43}
\end{equation}
\begin{equation}
D_{s}(s) = \int_{R_{0}}^{R_{1}} \frac{\partial}{\partial R}\,D \biggl ( \frac{sR}{R_{1}-R_{0}}, R \biggr )\,dR\, , \,\,\,\,{\rm for\,\,spherical\,\,wave\,\,propagation}\, ,
\label{eqp2v44}
\end{equation}
where the lower limit of integration $R_{0}$ is 
the radial distance from which scattering is assumed to be effective (we take this to be equal to the fundamental emission level), and the upper limit $R_{1}$ corresponds to the observer (here 
at $R_{1} =  1$ AU). All quantities are assumed to have 
spherical symmetry and the path is assumed to be radial.

Scattering depends sensitively on the ratio of the radiation 
frequency $f$ to the local electron plasma frequency 
$f_{p}(R)$ [Cairns, 1998]. Equations (16) and (22) of 
Cairns [1998] include the effects on refractive scattering 
that arise from $f_{p}(R)$ being non-zero and varying with 
position between the source and observer. By analogy with 
these equations we write
\begin{equation}
D_{pf}(s) = \int_{R_{0}}^{R_{1}} \frac{1}{1 - f_{p}(R)^{2}/f^{2}}\, \frac{\partial}{\partial R}\,D(s,R)\,dR
\label{eq3b2}
\end{equation}
for plane wave propagation and
\begin{equation}
D_{sf}(s) = \int_{R_{0}}^{R_{1}} \frac{1}{1 - f_{p}(R)^{2}/f^{2}}\, \frac{\partial}{\partial R}\,D \biggl ( \frac{sR}{R_{1}-R_{0}}, R \biggr )\,dR
\label{eq3b2a}
\end{equation}
for spherical wave propagation, respectively.

The scattering angle is conventionally defined using a coherence scale $s_{0}$ 
in the following manner [e.g., Coles et al., 1987; Bastian, 1994]:
\begin{equation}
\theta_{c} = \bigl ( 2 \pi s_{0}/\lambda \bigr )^{-1} \ ,
\label{eq3b2b} 
\end{equation}
where
\begin{equation}
D_{*}(s_{0}) = 1
\label{eq3b3}
\end{equation}
and $D_{*}(s)$ is either equal to $D_{pf}(s)$, defined by (\ref{eq3b2}), or 
$D_{sf}(s)$, defined by (\ref{eq3b2a}), in appropriate limits. This scattering 
angle $\theta_{c}$ can be interpreted as the 
predicted size of an idealized point source.

\subsection{Density Models}
A model for $n_{e}(R)$ in the corona and solar wind is required to be 
able to predict the angular broadening. The density model is required for 
computing the inner scale, which is defined in the next subsection.
Since there is no 
universally accepted model, we initially consider four representative density models. One 
is the four-fold Newkirk density model for the corona, based on eclipse 
observations [Newkirk, 1961]: 
\begin{equation}
n_{\rm 4n}(R) = 4 \times 4.2 \times 10^{4} \times 10^{4.32/R}\,\,\,\,\,\,\,\,\,{\rm cm^{-3}}\, ,
\label{eq3a1}
\end{equation}
The second model is derived from the frequency drift rate of interplanetary 
type III bursts [Leblanc et al., 1998]:
\begin{equation}
n_{\rm lb}(R) = 3.3 \times 10^{5}\,R^{-2} + 4.1 \times 10^{6}\,R^{-4} + 8 \times 10^{7}\,R^{-6}\,\,\,\,\,\,\,\,\,{\rm cm^{-3}}\, .
\label{eq3a2}
\end{equation}

The third model considered is due to Aschwanden et al. [1995]. It is 
based on the drift rates of type III bursts [Alvarez \& Haddock, 1973] 
in the outer corona  and solar  wind ($f < 10$ MHz) and assumes an 
isothermal barometric atmosphere for the lower corona:
\begin{equation}
n_{\rm a}(R) = \left\{ \begin{array}{ll} n_{1} \bigg (\frac{R-1}{R_{2}} \bigg)^{-p}\, , \,\,\,\,\,\,\,\,\,\,\,\,\,\,\,\,\,\,\,\,\, R > 1 + R_{2} \\
n_{Q} \exp \bigg(- \frac{R-1}{\mu} \bigg) \, , \,\,\,\,\,\,\, R < 1 + R_{2}
\end{array}
\right.
\label{eq3a3}
\end{equation}
where $p = 2.38$, $n_{Q} = 4.6 \times 10^{8}$ ${\rm cm}^{-3}$, $n_{1} = n_{Q} \exp (-p)$, 
$\mu = 0.1$ and $R_{2} = p \mu$.  
The fourth model is a ``hybrid'', using the Aschwanden \& Benz [1995] model 
for the lower corona and the four-fold Newkirk model multiplied 
by a normalization factor (to ensure continuity) in the upper corona. 
In other words, the density $n_{\rm hyb}(R)$ of the hybrid model is   
\begin{equation}
n_{\rm hyb}(R) = \left\{ \begin{array}{ll} A\,n_{\rm 4n}(R)\, , \,\,\,\,\,\,\,\,\,\,\,\,\,\,\,\,\,\,\,\,\, R > 1 + R_{2} \\
n_{\rm a}(R) \, , \,\,\,\,\,\,\, R < 1 + R_{2} \ , 
\end{array}
\right.
\label{eq3a4}
\end{equation}
where $A = 0.324$ is the normalization factor that ensures continuity.

\callout{Figure 2} shows $f_{p}(R) =  8.97\, n_{e}(R)^{1/2}$~kHz 
for all four density models, with $n_{e}$ in units of cm$^{-3}$. 

\begin{figure}
\noindent\includegraphics[width=40pc]{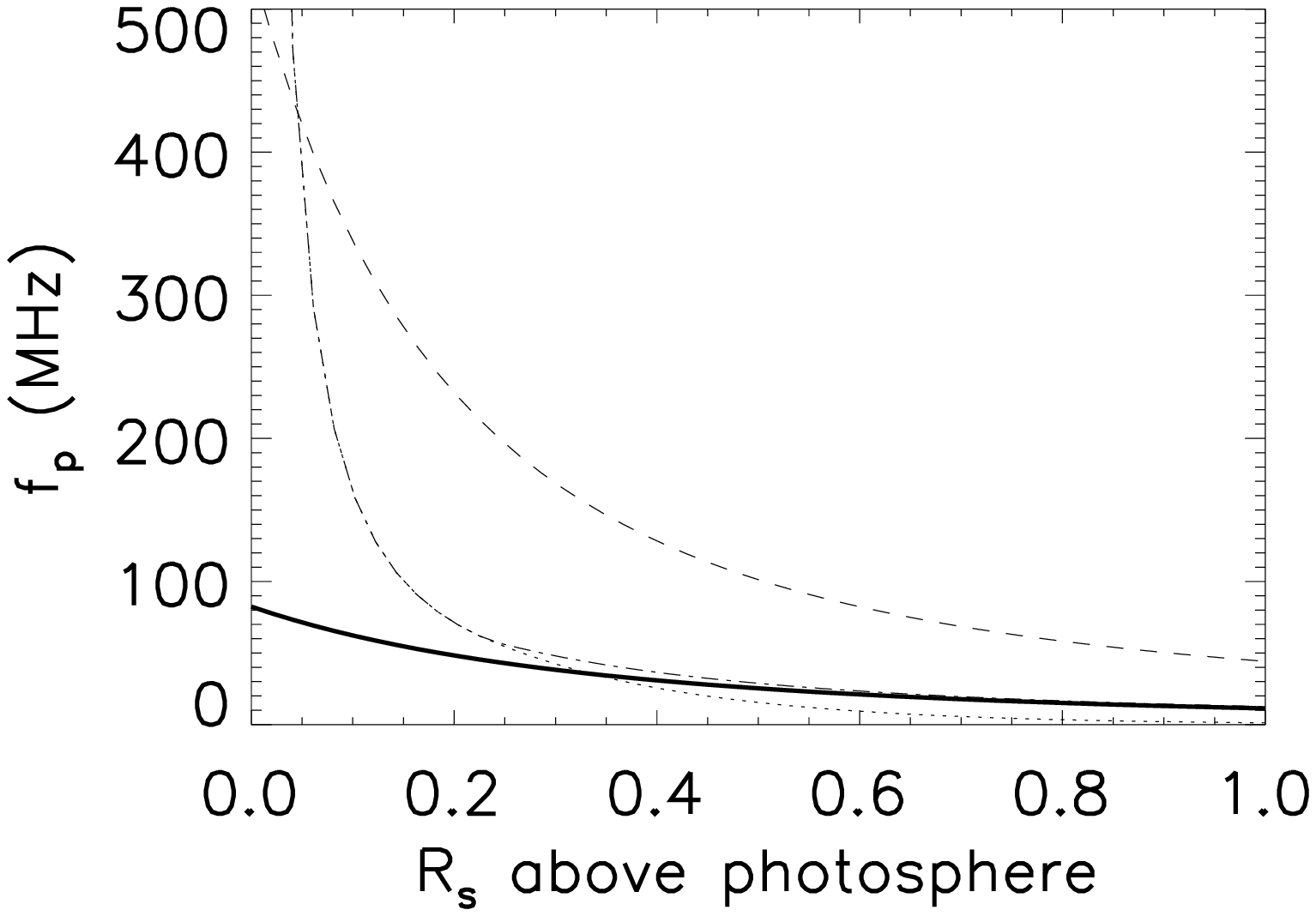}
\caption{The plasma frequencies predicted by the four 
density models in the text are plotted against the height 
$h = R - R_{\odot}$ above the photosphere (in units of 
$R_{\odot}$). The solid line uses the Leblanc et al model (\ref{eq3a2}), the
dotted line uses the Aschwanden \& Benz model (\ref{eq3a3}), the
dashed line uses the 4*Newkirk model (\ref{eq3a1}) and the
dash-dot line uses the ``hybrid'' model (\ref{eq3a4}).}
\end{figure}

The figure shows that the highest frequency predicted for $R > 1$ by the Leblanc 
et al. [1998] density model (Eq~\ref{eq3a2}) is less than 100 MHz. 
Since our observing frequency is 327 MHz, this model is therefore unsuitable 
for our purposes. However, the four-fold Newkirk model (\ref{eq3a1}), Aschwanden 
\& Benz [1995] model (\ref{eq3a3}), and hybrid model can account 
for $f_{p} = 327$ MHz for $R > 1$. However, since the Aschwanden \& Benz [1995] 
model predicts unrealistically low densities (and consequently $f_{p}$) for 
$R > 1 + R_{2} = 1.23$, only the hybrid model is 
considered further below. Fundamental emission at 327 MHz emanates from a 
heliocentric distance of 1.055 $R_{\odot}$ with this model. In order to 
avoid the singularity in the integrand in Eqs~(\ref{eq3b2}) and 
(\ref{eq3b2a}), we start the integration at $R_{0} = 1.056 \, R_{\odot}$. 
In other words, we start the integration from a distance of approximately 
700 km above the height at which 327 MHz fundamental emission originates. 
This distance is smaller than that corresponding to the frequency difference 
$\Delta f$ between the minimum frequency of fundamental emission at a given location 
and the local value of $f_{p}$, so that avoiding the singularity is 
correct. This positive frequency difference exists because  
conservation of energy and the wave dispersion relations force the 
standard nonlinear Langmuir wave processes for fundamental and harmonic 
emission to produce radiation of order several percent above $f_{p}$ 
and $2f_{p}$ [e.g., Cairns, 1987a,b], with the value of $\Delta f / f_{p}$ 
depending on the beam and plasma parameters. 

\subsection{Inner scale effects}

We next discuss the need for including inner scale effects in our treatment. In general, inner scale effects are important if the baseline is smaller than the inner scale. 

\begin{figure}
\noindent\includegraphics[width=40pc]{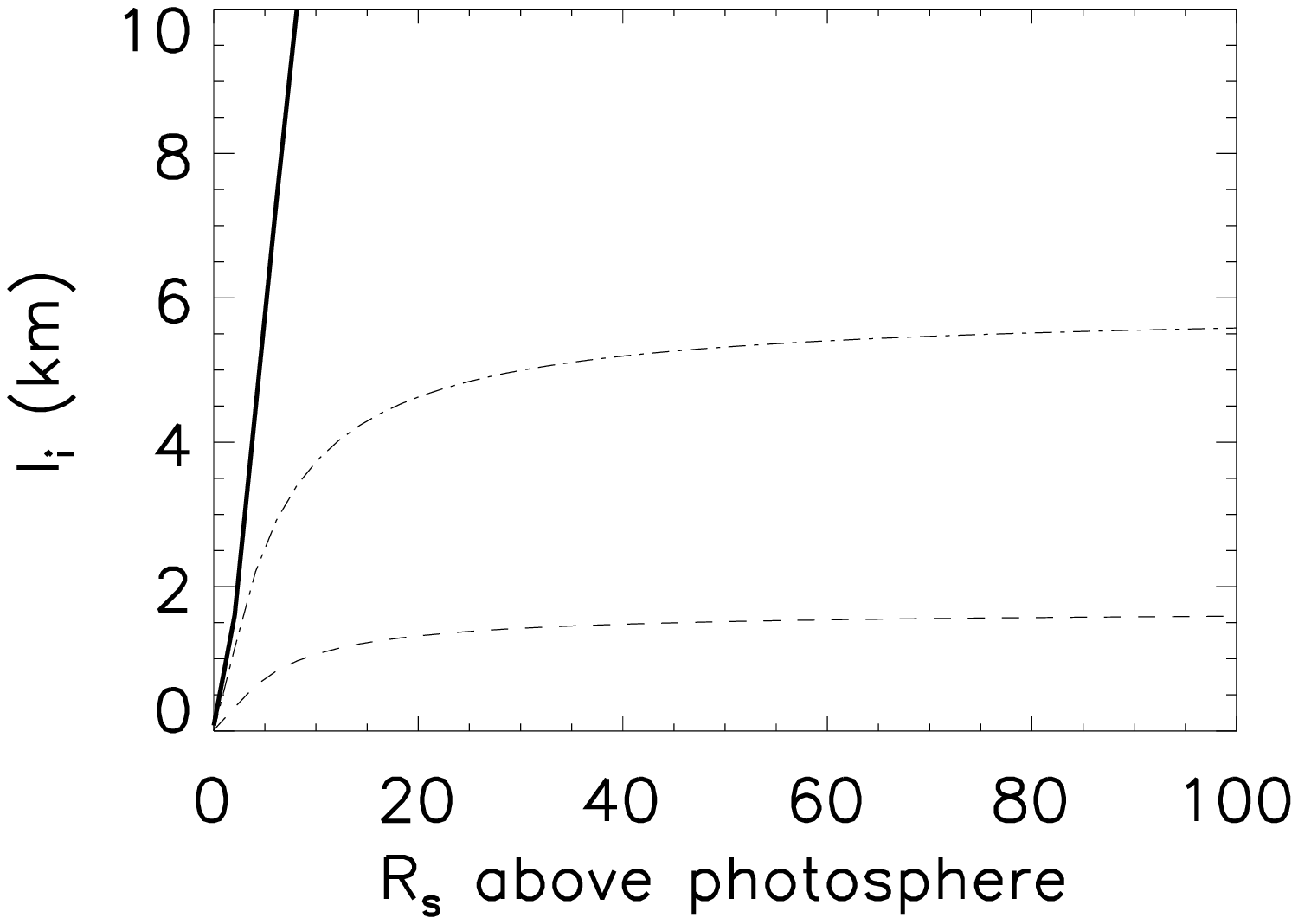}
\caption{The inner scales (in units of km) predicted by some 
density models in the text are plotted against the height 
$h = R - R_{\odot}$ above the photosphere (in units of $R_{\odot}$). 
The solid line uses the Leblanc et al model (\ref{eq3a2}), the
dashed line uses the 4*Newkirk model (\ref{eq3a1}), and the
dash-dot line uses the hybrid model (\ref{eq3a4}).}
\end{figure}

\callout{Figure 3} shows the inner scale (in km) given by Eq~(\ref{eq2}), as a function of heliocentric distance for some of the density models discussed in the preceding section. Clearly, the inner scale is quite dependent upon the density model. The inner scale for the Aschwanden \& Benz [1995] model far exceeds those for the other models, since the density with this model for $R > 1.23$ in unrealistically low; we have therefore chosen not to depict the $l_{i}$ for this model in Figure 3. As explained in the preceding section, we choose to use only the hybrid model from now on, for it is by far the most realistic one.

In order to ascertain the importance of inner scale effects, we compare $l_{i}$ with the longest baseline (that determines the smallest source size), assuming that the source is situated at the fundamental emission level for 327 MHz, and the observer is at 1 AU. In order to compute the longest baseline, we set $49^{''} = 1.22 \lambda/s$, where $49^{''}$ is the observed source size and $\lambda$ is the (free space) observing wavelength (1 meter). This yields an effective baseline of $s \sim 5$ km. In order to ascertain the relevance of the inner scale (i.e., whether we should be using Eq~\ref{eqp2v41} or \ref{eqp2v42}) we need to compare the effective longest baseline $s_{\rm eff}$ with the 
inner scale. It may be noted that a typical interferometer measurement involves a range of baselines, and while the longest baseline we have computed above is the one that limits the smallest observable source size, baselines shorter than this one do contribute to the overall measurement. 
However, our approach is appropriate because the longest baseline 
for a given source size is the largest length scale at which there is appreciable 
power. Inner scale effects are relevant only for the branch for which 
the baseline is $\ll$ the inner scale, as in Eq~\ref{eqp2v42}. If the {\em longest} 
relevant baseline is smaller than the inner scale, then it follows that 
the rest of the baselines in the problem automatically satisfy this 
criterion. Our approach thus provides a useful estimate of the importance 
of inner scale effects.

As discussed above, $s_{\rm eff} = s$ for plane wave propagation, 
while $s_{\rm eff} = s R/(R_{1} - R_{0})$ for spherical wave propagation. We show 
the ratio of $s_{\rm eff}$ to $l_{i}$ in \callout{Figure 4}.

\begin{figure}
\includegraphics[width=40pc]{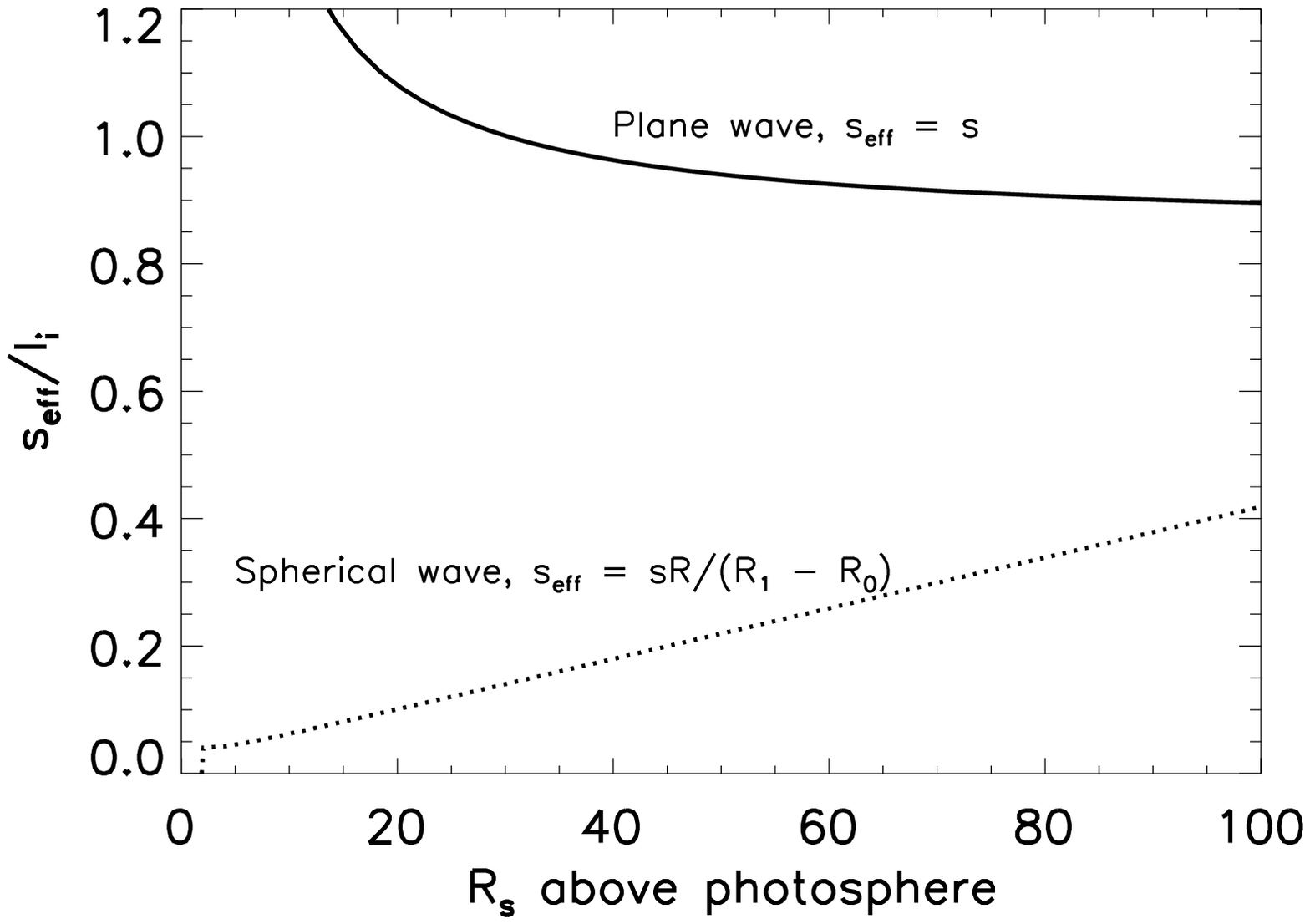}
\caption{The ratio $s_{\rm eff}/l_{i}$ is plotted against 
the height $h = R - R_{\odot}$ above the photosphere (in 
units of $R_{\odot}$). The solid line shows the ratio for 
plane wave propagation (i.e., with $s_{\rm eff} = s$), while 
the dotted line shows the ratio for spherical wave propagation 
(i.e., with $s_{\rm eff} = s R/(R_{1} - R_{0})$). The hybrid 
density profile is used for determining the density and the inner scale.}
\end{figure}

When considering plane wave propagation, $s_{\rm eff} = s$, and the solid line in Figure 4 shows that $s_{\rm eff}$ is mostly $> l_{i}$. For heliocentric distances greater than about 30 $R_{\odot}$, $s_{\rm eff}$ does become somewhat smaller than $l_{i}$, but most of the contribution to the scattering kernel arises from distances well inside 30 $R_{\odot}$. We should therefore use Eqs~(\ref{eqp2v41}) and (\ref{eq3b2}) for plane wave propagation. On the other hand, when considering spherical wave propagation, $s_{\rm eff} = s R/(R_{1} - R_{0})$, and the dotted line in Figure 4 shows that $s_{\rm eff}$ is $< l_{i}$ for all $R$. The appropriate equations to use for spherical wave propagation are therefore Eqs~(\ref{eqp2v42}) and (\ref{eq3b2a}).






\section{Results}
\subsection{Plane wave propagation}
We first consider plane wave propagation, which is more appropriate for waves emanating from a background object that is far from the scattering medium.
\subsubsection{$s_{\rm eff} \gg l_{i}$}

Since Figure 4 shows that $s_{\rm eff} > l_{i}$ for plane wave propagation 
for $R < 30 R_{\odot}$, the appropriate branch to use is Eq~(\ref{eqp2v41}). 
At heliocentric distances greater than about 30 $R_{\odot}$, $s_{\rm eff}$ 
becomes marginally less than the inner scale, but we have verified numerically 
that this is immaterial, since most of the contribution to $\theta_{c}$ takes 
place well within 30 $R_{\odot}$.

\begin{figure}
\noindent\includegraphics[width=40pc]{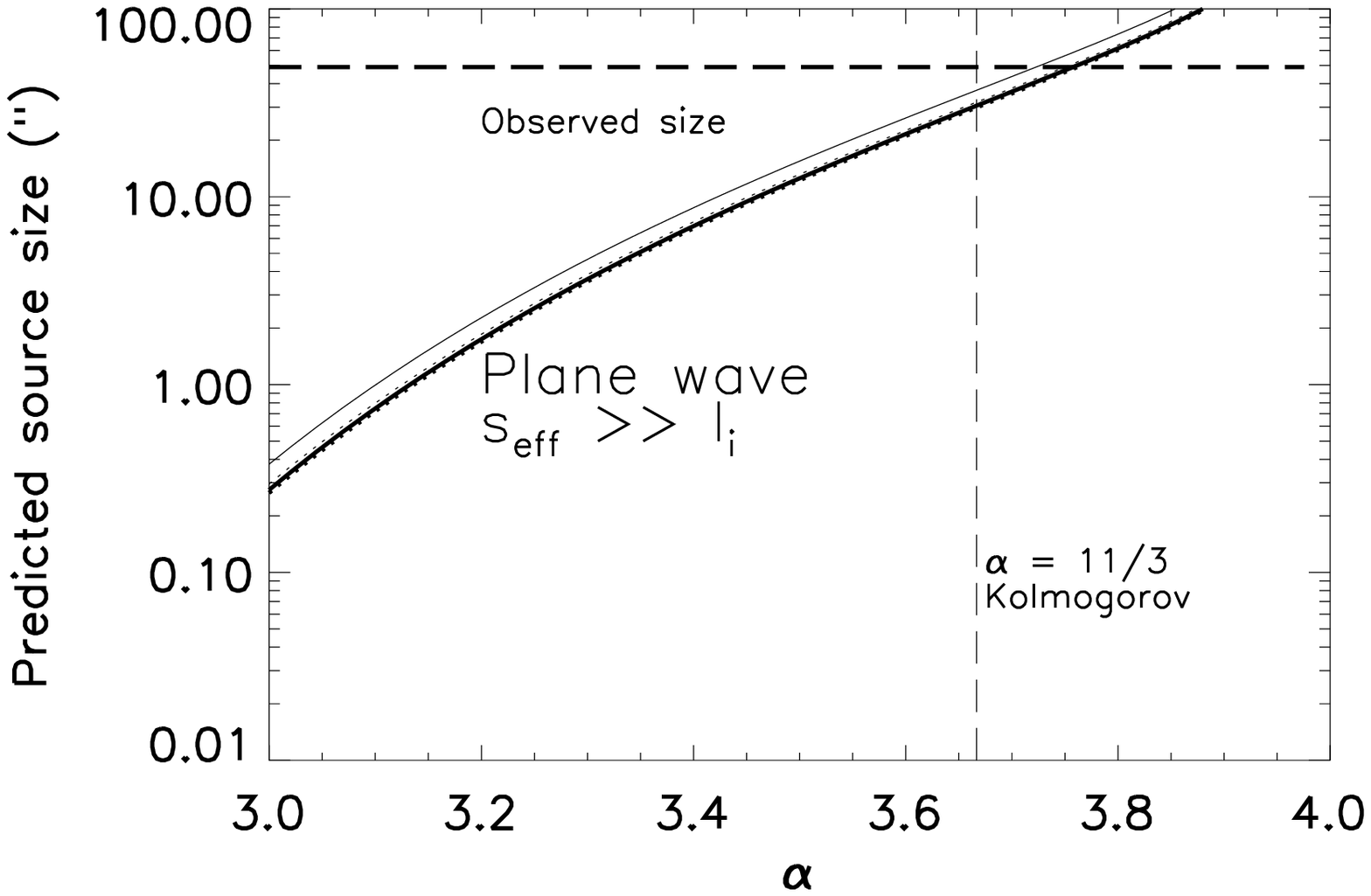}
\caption{Predicted $\theta_{c}$ (1 AU) in $^{''}$ at 327 MHz for 
plane wave propagation, as a function of $\alpha$, using 
(\ref{eqp2v41}), (\ref{eq3b2}) and (\ref{eq3b3}). The observed 
source size is 49$^{''}$. The thin line is for fundamental emission 
and the thick line is for second harmonic emission.}
\end{figure}

\callout{Figure 5} uses Eqs (\ref{eqp2v41}), 
(\ref{eq3b2}) and (\ref{eq3b3}) to predict the scattering angle $\theta_{c}$(1 AU) 
at the Earth, in arcseconds, as a function of the power law index $\alpha$ for 
plane wave propagation at $f = 327$~MHz. The thin line is for fundamental emission 
and the thick line is for second harmonic emission. Removal of the refractive 
index effect in Eq~(\ref{eq3b2}), meaning the factor of $(1 - f_{p}^{2}/f^{2})^{-1}$), 
causes a negligible change in the result. The predicted source size is slightly smaller 
for second harmonic emission.

\subsubsection{$s_{\rm eff} \ll l_{i}$}

Although Figure 4 demonstrates that $s_{\rm eff} > l_{i}$ for plane wave propagation, 
we nevertheless investigate the predicted scattering angle for plane wave propagation, 
while using branch (\ref{eqp2v42}). The results are shown in \callout{Figure 6}. 

\begin{figure}
\noindent\includegraphics[width=40pc]{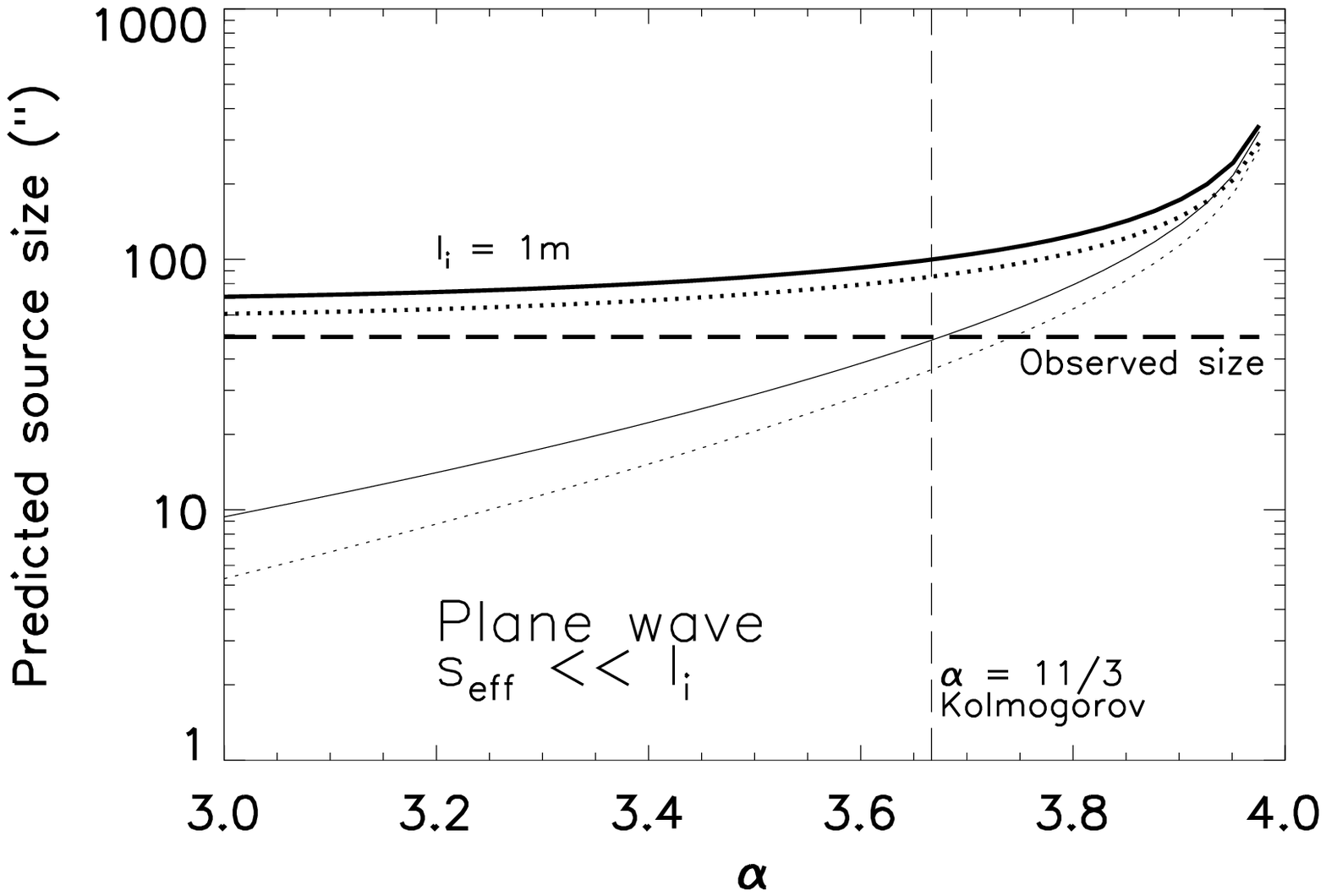}
\caption{Predicted $\theta_{c}$ (1 AU) in $^{''}$ at 327 MHz for plane wave propagation as a function of $\alpha$, using Eqs (\ref{eqp2v42}), (\ref{eq3b2}) and (\ref{eq3b3}). The solid line is for fundamental emission and the dotted line is for second harmonic emission. The heavy lines are computed with $l_{i}$ set at an artificially low value of 1 m, while still using branch (\ref{eqp2v42}).}
\end{figure}

Evidently, the source sizes predicted approach the observed size for relatively steep turbulence spectra; for spectra that are steeper than Kolmogorov (i.e., $\alpha > 11/3$), the predicted source sizes exceed the observed one. 
In order to investigate inner scale effects we compute the scattering angle for 
plane wave propagation with the inner scale set to an artificially low value of $1$~m, 
instead of being computed self-consistently from Eq~\ref{eq2}. The results are shown 
using the heavy lines in \callout{Figure 7}. When inner scale effects are thus removed, it is 
clear that the predicted source size increases, especially for flatter spectra. 

\subsection{Spherical wave propagation}
As discussed earlier, spherical wave propagation is appropriate when the source is embedded in the scattering medium, as is the case here.  

\subsubsection{$s_{\rm eff} \ll l_{i}$}

Since Figure 4 shows that $s_{\rm eff} \ll l_{i}$ for spherical wave propagation, the appropriate branch to use is (\ref{eqp2v42}). 

\begin{figure}
\noindent\includegraphics[width=40pc]{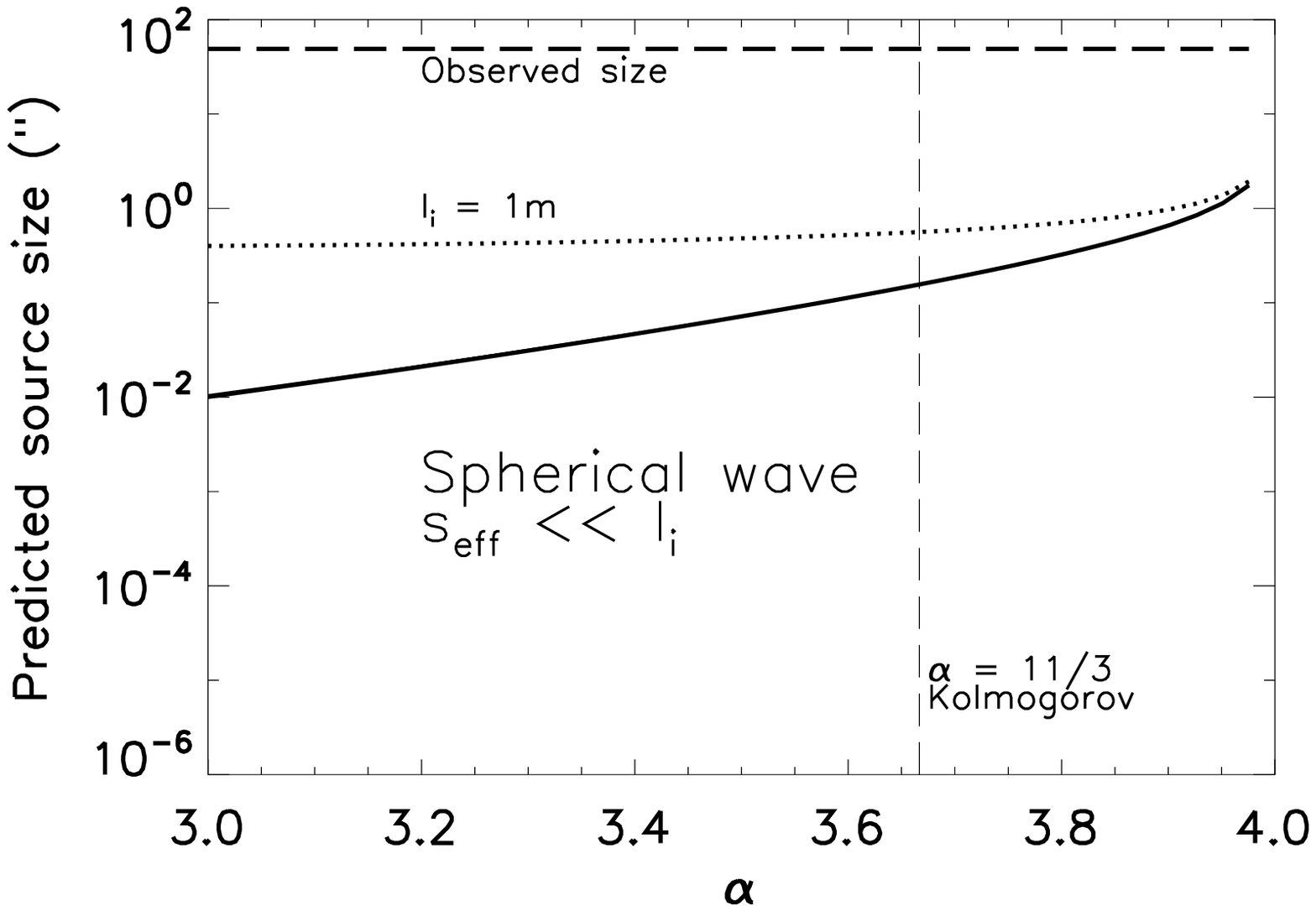}
\caption{Predicted $\theta_{c}$ (1 AU) in $^{''}$ at 327 MHz for spherical wave propagation as a function of $\alpha$, using (\ref{eqp2v42}), (\ref{eq3b2a}) and (\ref{eq3b3}). The observed source size at 327 MHz is 49$^{''}$. There is negligible difference between fundamental and second harmonic emission. The solid lines are computed with $l_{i}$ from prescription (\ref{eq2}), while the dotted lines are computed with $l_{i}$ set to an artificially low value of 1 m, while still using branch (\ref{eqp2v42}).}
\end{figure}

The solid line in Figure 7 predicts $\theta_{c}$(1 AU) for spherical wave propagation, using (\ref{eqp2v42}), (\ref{eq3b2}) and (\ref{eq3b3}). Clearly, the predicted scattering angle is at least 25 times smaller than the observed one. The dashed line, on the other hand, is computed by artificially setting $l_{i} = 1$m, while still using (\ref{eqp2v42}). This is tantamount to neglecting inner scale effects. We observe that for flat spectra, inner scale effects substantially reduce (by over an order of magnitude) the predicted $\theta_{c}$, but that this difference is progressively reduced as $\alpha$ increases. There are negligible differences between the results for fundamental and second harmonic emission, and removal of the refractive index effect causes a negligible change too.

\subsubsection{$s_{\rm eff} \gg l_{i}$}
In keeping with the spirit of our treatment for plane wave propagation, we investigate the predicted scattering angle for spherical wave propagation while using branch (\ref{eqp2v41}), which assumes that $s_{\rm eff} > l_{i}$. We do this despite Figure 4's prediction that $s_{\rm eff}$ is  $< l_{i}$ for spherical wave propagation. 

\begin{figure}
\noindent\includegraphics[width=40pc]{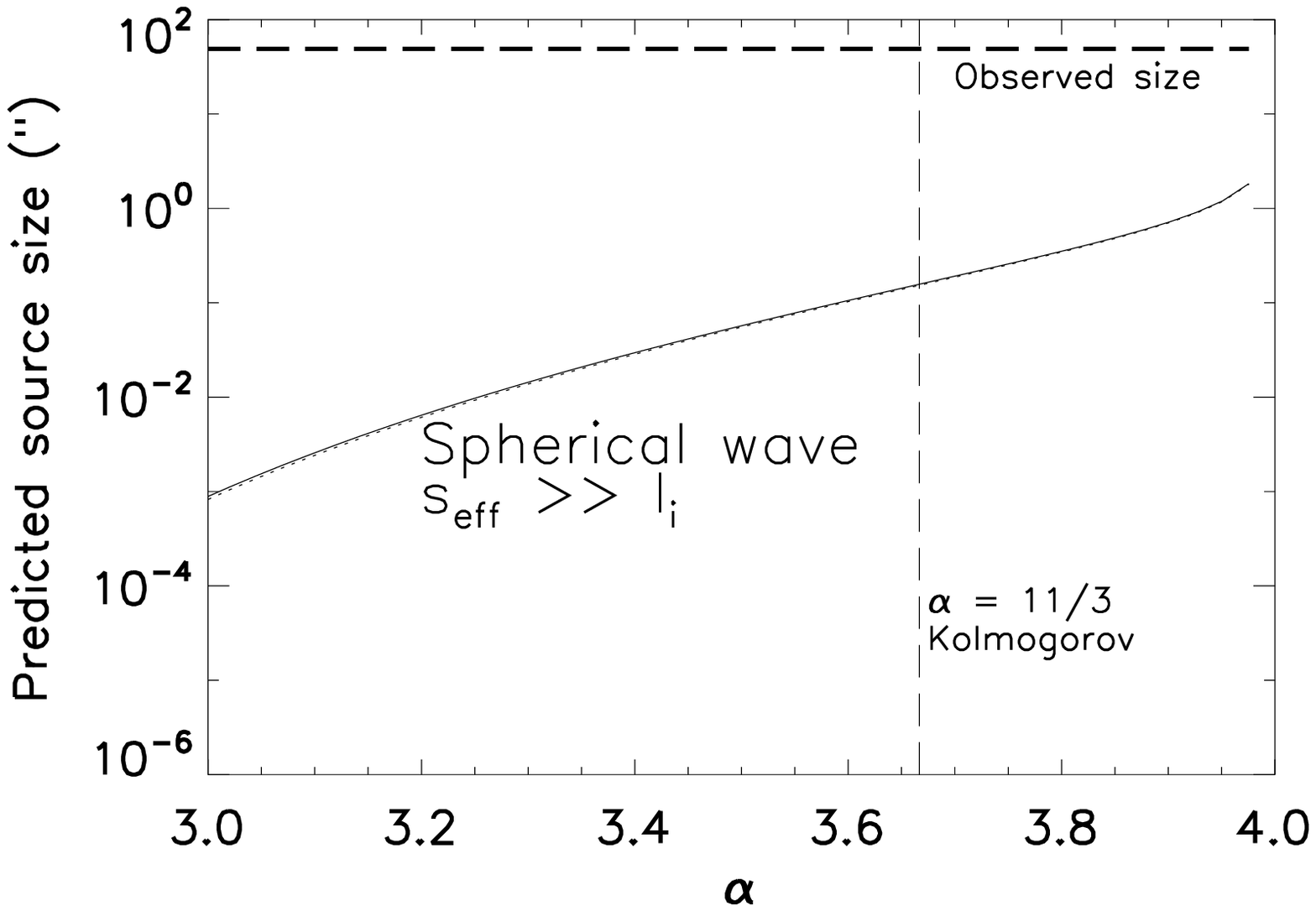}
\caption{Predicted $\theta_{c}$ (1 AU) in $^{''}$ at 327 MHz for spherical wave propagation as a function of $\alpha$, using (\ref{eqp2v41}), (\ref{eq3b2a}) and (\ref{eq3b3}).. The difference between the predictions for fundamental and second harmonic emission is negligible.}
\end{figure}
The difference between fundamental and second harmonic emission are negligible.

\subsubsection{Spherical vs plane wave propagation}
Although we have investigated several different cases, our attention has been focussed mainly on two issues: first, the difference between the source sizes predicted for plane wave and spherical wave propagation, and, second, the influence of the inner scale. We now compare the plane wave and spherical wave results directly in \callout{Figure 9}, assuming fundamental emission, and employing an inner scale that is computed self-consistently using Eq~(\ref{eq2}).

\begin{figure}
\noindent\includegraphics[width=40pc]{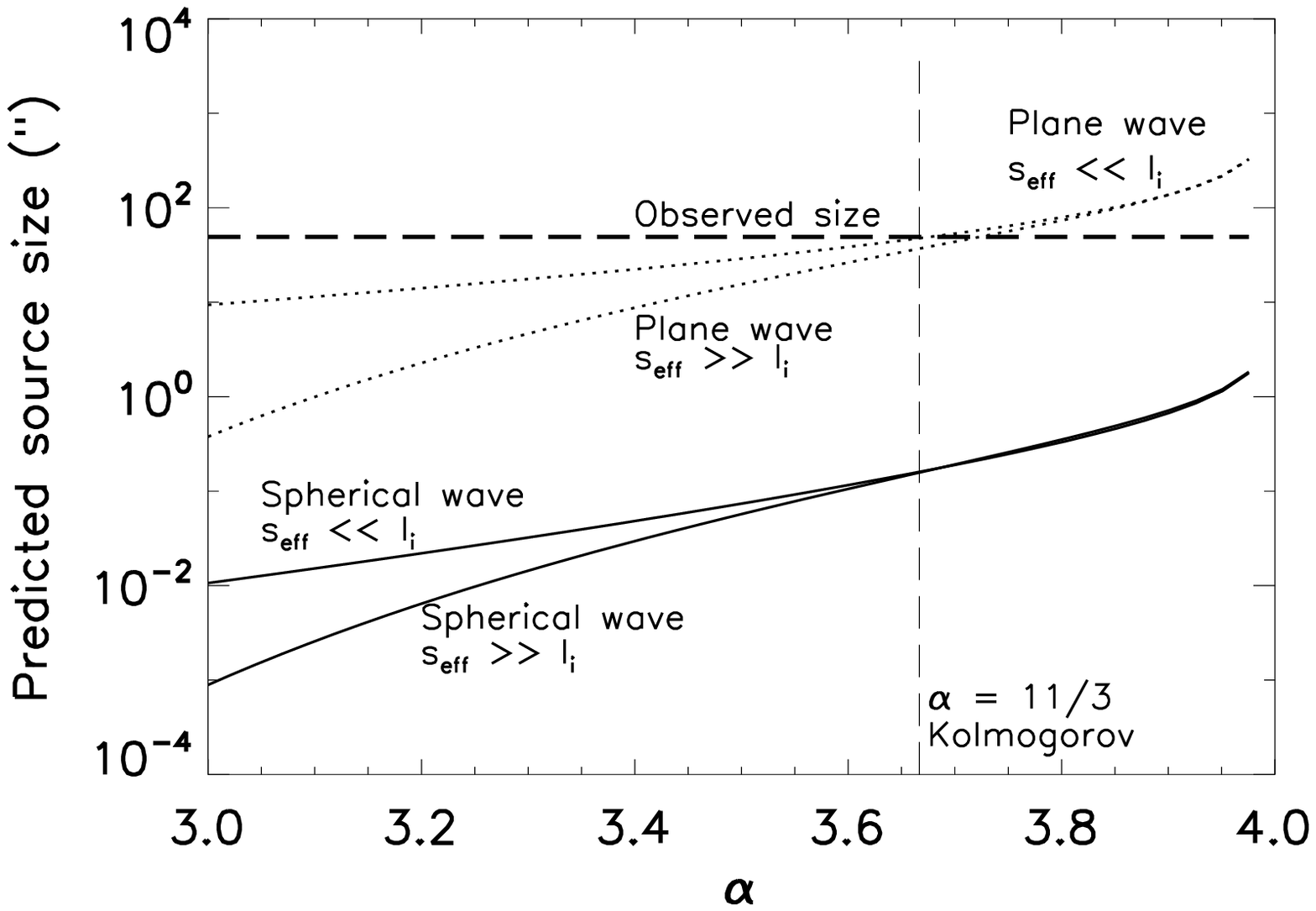}
\caption{Direct comparison of $\theta_{c}$ (1 AU) for spherical (solid lines) and plane wave (dotted lines) propagation. Fundamental emision is assumed.}
\end{figure}

It is clearly evident from Figure 9 that spherical divergence effects decrease the predicted scattering angle by around two orders of magnitude as compared to the plane wave case. This is best seen by comparing the same branch (say, $s_{\rm eff} > l_{i}$) for the plane wave and spherical wave cases. Thus, spherical divergence effects should be quantitatively important for scattered solar radio emission and plane wave results should be used with great caution.

\section{Discussion and Summary}
The highest resolution meter wavelength observations of the solar 
corona reveal compact sources around 49$^{''}$ in size at 327 MHz. 
The main aim of this paper is to employ an observationally-motivated 
model for the turbulence amplitude $C_{N}^{2}$, and see what it 
implies for the predicted scattering angle for radio sources located 
in the solar corona. We reference our calculations to the same 
frequency (viz. 327 MHz) at which the smallest source size is observed. 
We employ the parabolic wave equation, together with the standard 
asymptotic forms for the phase structure function, which are valid 
for situations where the effective baseline is either much larger 
or much smaller than the inner scale. We define the predicted scattering 
angle $\theta_{c}$ via Eq. (\ref{eq3b2b}) as the angle where 
the phase structure function falls to 1/e times 
its peak value. Effectively, this means that the  
scattering angles predicted here should be interpreted as the scatter-broadened 
image of an ideal point source in the solar corona. The real source will 
have an intrinsic size (i.e., it will not be a point source) and the 
observable source will be the convolution of the intrinsic source 
profile with $\theta_{c}$, 
provided there are no instrumental limitations. The results in this paper 
should therefore be regarded as lower limits to the observable source 
size set by scattering. The general consensus now seems to be that 
there is not much about the instrinsic source size that can be gleaned 
from scatter-broadened images [Bougeret \& Steinberg, 1977; Melrose, 1980; Bastian, 1994]. However, 
if the intrinsic source size and $\theta_{c}$ are similar in size, then the 
observed source size will be larger than $\theta_{c}$ by a factor 
near $\sqrt{2}$. Note that this factor cannot account for the large 
discrepancies between the minimum source size observed at 327 MHz [Mercier 
et al., 2006] and those predicted here. If, on the other hand, the 
intrinsic source size is much larger, then scatter broadening does not play 
a significant role. 

We have included refractive index effects that can be important when 
the radiation is emitted near the fundamental plasma level, but found them to be 
relatively unimportant. The inner scale is included via the Coles and Harmon 
[1986] model, interpreted in terms of cyclotron damping of MHD waves, and 
so depends primarily on the ambient electron density. We employ a hybrid 
model for the electron density that yields reasonable heights for meter 
wavelength emission at the fundamental. In view of the 
uncertainity in its value in the inner corona, the power law index 
$\alpha$ characterizing the turbulent spectrum is taken to be a free 
parameter. We consider both plane wave and spherical wave propagation. 
For the geometry we consider, where the source is embedded in the scattering 
medium, the spherical wave description is arguably more appropriate.

We have thus explored a wide variety of effects. We observe that 
there is no significant difference in the predicted scattering angle 
between fundamental and second harmonic emission. We also find that 
the removal of the refractive index effect causes a negligible change 
in the predicted scattering angle. We find that the spherical 
divergence effect results in a significant lowering of the predicted 
scattering angle (by around 2 orders of magnitude). We find that 
removing inner scale effects by artificially setting the inner scale 
to be equal to a very small value (instead of determining it 
self-consistently from Eq~\ref{eq2}) results in a significant 
enhancement of the predicted source size. The enhancement is 
greatest for flatter spectra, where it can be a factor of 
around $50$, and it progressively disappears for steeper spectra.

As mentioned earlier, the power law index of the turbulent spectrum 
is a free parameter. There is a formal divergence at $\alpha = 4$ 
in Eqs (\ref{eqp2v41}) and (\ref{eqp2v42}), and we therefore limit the 
computations to a maximum value of $\alpha = 3.97$. The maximum value 
of the predicted scattering angle thus occurs at $\alpha = 3.97$.

For plane wave propagation, the predicted source size for Kolmogorov turbulence is around 10$^{''}$ lower than the observed one. For spherical wave 
propagation, we find that the maximum value of the predicted scattering 
angle is at least 25 times smaller than the observed one. It is emphasized that plane wave propagation is relevant to the 
well-accepted empirical formula of Erickson [1964], which predicts large 
source sizes, since it pertains to observations of celestial sources 
through the solar wind. Even so, with current estimates of $l_{i}(R)$ 
implying that $s_{\rm eff} > l_{i}$ (Figure 4), 
additional scattering is required to bring Erickson's result into 
quantitative agreement with the calculations here for plane wave 
scattering. Alternatively, the inner scale $l_{i}(R)$ should be 
smaller than that predicted by the Coles and Harmon [1989] 
model assumed here, so that $s_{\rm eff} < l_{i}$.   

The crucial result of this paper is that the 
predicted source sizes are considerably smaller than the observed 
lower limit of $49''$ when spherical wave propagation and 
inner scale effects are 
included, as they should be for sources in the solar corona. 
This broad trend of the models substantially underpredicting the 
source size can be interpreted in three ways that are not exclusive 
and can occur in combination. First, it could imply 
that source sizes much smaller than those that have been observed 
so far actually exist in the solar corona, and can potentially be 
observed. All the instances of observations of small sources to date 
have been limited by the instrument resolution; it is therefore quite 
likely that smaller sources can be detected when instrument resolutions 
are improved. Second, this broad trend can be taken to imply that 
our naive extrapolation of the empirical form  for the 
turbulence amplitude $C_{N}^{2}(R)$  to the inner corona is not 
justified. The results could be taken to imply that $C_{N}^{2}(R)$ in the 
inner corona is far higher than suggested by the empirical formula 
(\ref{eq3}). This could be due to the functional 
form for $C_{N}^{2}(R)$ increasing more rapidly with decreasing $R$ or 
due to a larger normalization 
factor or both effects. Third, the model (\ref{eq2}) may significantly 
overestimate $l_{i}(R)$, meaning that the turbulent cascade extends 
to smaller length scales (larger $q$) and leads to more 
scattering. These proposals all appear reasonable, and we regard all 
three as viable.

Finally, we discuss the connection between our work and that of 
Bastian [1994]. The methodology is similar, and we investigate similar 
issues such as the effects of the inner scale, turbulence index $\alpha$, and 
spherical versus planar wave propagation. Bastian's [1994] findings 
are contrary to ours: we find that our model predictions are substantially 
below the minimum observed size of $49''$, while Bastian's [1994] model 
predictions are substantially above $49''$. Thus, a priori, both 
models need revision. 
A major difference is in the choice of a model for $C_{N}^{2}$. 
Bastian [1994] uses a model for $C_{N}^{2}$ which
is proportional to the square of the background electron 
density and assumes the density model of Riddle (1974), which 
also involves a constant of proportionality. These two constants of 
proportionality are absorbed into one and fixed by normalizing 
the structure function $D_{20}(10 {\rm km})$ for a baseline of 10 km, an observing wavelength 
of $20$ cm, and an elongation of 5 $R_{\odot}$ . In contrast, as explained 
earlier, the $C_{N}^{2}$ model we use is determined by an empirical fit to 
VLBI scattering observations between 10--50 $R_{\odot}$; this was motivated by the 
need to use a $C_{N}^{2}$ model that is derived as directly as possible from 
observations. A minor matter is that Bastian [1994] discusses the 
disk to limb variation in the predicted scattering angle, whereas our treatment 
is valid only for sources that are reasonably close to disk center. In order to do so, we would need to use the general formalism used here, together with an integration path that incorporates the appropriate extra path length needed for sources that are displaced from the disk center.

An appropriate means of 
comparing the normalizations of the two treatments is 
thus to compare the normalizations of the structure function. 
Using (\ref{eq3b2}) -- (\ref{eq3b3}), we write the structure 
functions as 
{\begin{eqnarray}
\nonumber
D_{sf}(s) = \frac{4 \, \pi^{2}\,s^{2}}{\lambda^{2}}\,\theta_{c\,sf}^{2} \, , \\
D_{pf}(s) = \biggl ( \frac{2\, \pi\, s}{\lambda} \biggr )^{\alpha - 2}\, \theta_{c\,pf}^{\alpha - 2}\, ,
\label{eq3a5}
\end{eqnarray}}
where $\theta_{c\, sf}(s)$ is the value of $\theta_{c}$ for spherical wave 
propagation using branch (\ref{eqp2v42}) and $\theta_{c\, pf}(s)$ 
corresponds to plane wave propagation for branch (\ref{eqp2v41}). Then using 
our models for $C_{N}^{2}(R)$, $l_{i}(R)$, and $n_{hyb}(R)$ we find that 
$D_{sf}(s) = 2.8 \times 10^{-3}$ rad$^{2}$ for s = 10 km, $\lambda$ = 91 cm 
(corresponding to 327 MHz), $\alpha = 11/3$ and a starting height corresponding 
to 327 MHz fundamental emission. The same prescription and parameters yield 
$D_{pf}(s) = 7.7 \times 10^{-3}$ rad$^{5/3}$. Since 
the structure functions we derive are based on integrations 
over heliocentric distance, we cannot assign a specific elongation to them. 

In comparison, Bastian normalizes $C_{N}^{2}$ by assuming 
$D_{20 {\rm cm}}(10 {\rm km}) = $4--12 rad$^{2}$, based on measurements of 
by Coles and Harmon [1989] and Armstrong et al. [1990] of cosmic sources (implying primarily 
planar wave effects) 
at an elongation of 5 $R_{\odot}$. In order to normalize 
Bastian's [1994] values for the structure function to a wavelength of 91 cm, 
we concentrate on the structure function for spherical wave propagation. 
Inspection of Eqs [\ref{eq3b2a}] and [\ref{eq3a5}] reveals that $D_{sf}(s) 
\propto \lambda^{2}$. Therefore, $D_{20 {\rm cm}}(10 {\rm km})/D_{91 {\rm cm}}(10 {\rm km}) 
= 21$. Bastian's [1994] range of values for $D_{20 {\rm cm}}(10 {\rm km})$ thus 
corresponds to $D_{91 {\rm cm}}(10 {\rm km}) = 82 - 250$~rad$^{2}$. 
The difference in $D_{91 {\rm cm}}(10 {\rm km})$ between the two prescriptions 
is thus a factor of $\approx (30 - 90)\times 10^{3}$, corresponding to a factor 
$\approx 170 - 300$ in $\theta_{c}$. 

This large difference $\approx (30 - 90)\times 10^{3}$ in the normalization 
of the structure function is primarily indicative of a corresponding difference 
in the normalization of $C_{N}^{2}$ between the two treatments; this is 
because neglect of inner scale effects increases $\theta_{c}$ by 
less than a factor of $10$ for Kolmogorov turbulence in Figure 7. In this connection, we
note that Bastian's [1994] normalization for $D_{20 {\rm cm}}(10 {\rm km})$ 
is based on values of $D(s)$ measured at an elongation 
of 5 $R_{\odot}$. We also note (e.g., Fig 1 of Coles \& Harmon [1989]) that values 
of $D(s)$ measured at larger elongations can be considerably lower (by as much 
as a few orders or magnitude, depending upon the elongation). This is significant, since the model for $C_{N}^{2}$ that we use in this paper is based on observations between 10 and 50 $R_{\odot}$. 

In summary, the foregoing results demonstrate conclusively that 
spherical wave propagation effects are vital for solar sources, with 
plane wave predictions several orders of magnitude larger than the 
spherical predictions. Similarly, inner scale effects are quantitatively 
important, while fundamental versus harmonic radiation effects are relatively 
small. The results and discussion above demonstrate the importance of 
accurate models for $C_{N}^{2}(R)$ and to a lesser extent models of 
$l_{i}(R)$ and $\alpha(R)$. This paper's prescription for $C_{N}^{2}$ (Eq~[\ref{eq3}]) is 
empirical and directly based on observations (but extrapolated 
to smaller $R$), does not have any normalization 
constants that need to be determined, and leads to scattered sizes 
for a point source that are smaller than the minimum source size 
observed to date ($49''$ by Mercier et al. [2006]). Thus smaller 
source sizes than $49''$ may be observable. In contrast, another well-known prescription [Bastian, 1994] 
predicts much stronger scattering with source sizes always larger than $49''$: 
while this is inconsistent with the minimum source size observed to date 
at 327 MHz, it may provide the extra scattering required to account for 
Erickson's empirical angular broadening result for cosmic sources viewed 
through the solar wind. 

While this paper's results extend and confirm 
previous theoretical results pertaining to spherical vs. plane wave effects and provide the first explanation of the 
small source sizes recently 
observed, it is also clear that more observational and theoretical work 
is required on $C_{N}^{2}(R)$ especially, but also on $l_{i}(R)$ and $\alpha(k,R)$. 
This includes temporal variations over the solar cycle but also spatial 
variations between radio source 
regions and other regions of the corona. Increases in $C_{N}^{2}(R)$ and 
decreases in $l_{i}(R)$ would lead to more scattering. Work on both $n_{e}(R)$ and 
$\delta n(R) / n_{e}(R)$ may be useful 
[e.g., Efimov et al., 2008; Cairns et al., 2009]. It is quite possible that 
scattering observations and theory will provide useful constraints on 
these five quantities and therefore on the processes heating the solar corona and 
accelerating the solar wind.





%








%




%




%




%




%





%


%






%








%


\begin{acknowledgments}

PS acknowledges several illuminating discussions with Prof Rajaram Nityananda. The authors acknowledge several critical observations made by the anonymous referees, which have improved the contents of this paper.
PS acknowledges support from the Endeavour India research fellowship administered 
by the Department of Education, Science and Technology, Australian Government, which 
supported his visit to the University of Sydney. The initial part of this work was carried out when he was at his previous position at the Indian Institute of Astrophysics. IHC acknowledges the support of 
the Australian Research Council.

\end{acknowledgments}

\end{article}






%






%




%



%





%




%



\end{document}